\documentclass[twocolumn,dvipsnames]{aastex62}

\usepackage{CJK}
\usepackage{framed,amsmath,bm}

\graphicspath{{./}{figures/}}

\submitjournal{ApJ}

\shorttitle{}
\shortauthors{Zhu \& Gnedin}

\begin{document}
\begin{CJK*}{UTF8}{gkai}

\title{On the Properties of Cosmological Ionization Fronts}

\correspondingauthor{Hanjue Zhu (朱涵珏)}
\email{hanjuezhu@uchicago.edu}
\author[0000-0003-0861-0922]{Hanjue Zhu (朱涵珏)}
\affiliation{Department of Astronomy \& Astrophysics; 
The University of Chicago; 
Chicago, IL 60637, USA}

\author[0000-0001-5925-4580]{Nickolay Y.\ Gnedin}
\affiliation{Theory Division; 
Fermi National Accelerator Laboratory;
Batavia, IL 60510, USA}
\affiliation{Kavli Institute for Cosmological Physics;
The University of Chicago;
Chicago, IL 60637, USA}
\affiliation{Department of Astronomy \& Astrophysics; 
The University of Chicago; 
Chicago, IL 60637, USA}

\begin{abstract}
We investigate the properties of cosmological ionization fronts during the Epoch of Reionization using the CROC simulations. By analyzing reionization timing maps, we characterize ionization front velocities and curvatures and their dependence on the density structure of the intergalactic medium (IGM).  The velocity distribution of ionization fronts in the simulations indicates that while the barrier-crossing analytical model captures the overall shape in high-velocity regions, it fails to reproduce the low-velocity tail, highlighting the non-Gaussian nature of the IGM's density field. Ionization front velocities are inversely correlated with local density, propagating faster in underdense regions and more slowly in overdense environments. Faster ionization fronts also lead to higher post-ionization temperatures, reaching a plateau at $\sim 2 \times 10^4$ K for velocities exceeding 3000 km/s. Examining curvature statistics further establishes a connection between ionization front structure and the normalized density contrast $\nu$, with trends in overdense regions aligning well with barrier-crossing model predictions, while deviations appear in underdense environments due to model limitations. These results provide a detailed characterization of ionization front dynamics and their interaction with the underlying density field, bridging small-scale reionization physics with large-scale observables such as the 21 cm signal and the IGM's thermal history. \\

\end{abstract}

%\keywords{galaxies --- methods, numerical --- cosmology}

%%%%%%%%%%%%%%%%%%%%%%%%%%%%%%%%%%%%%%%
\section{Introduction}\label{sec:intro}

Cosmic reionization represents a critical epoch in the universe's history when the intergalactic medium (IGM) transitioned from being largely neutral to nearly fully ionized. This transition, driven by ultraviolet photons from the first stars, galaxies, and quasars, set the thermal and ionization states of the IGM and created the Lyman-alpha forest we observe at lower redshifts \citep{Shapiro&Giroux1987,Madau1999,Finkelstein2015,Robertson2015,Dayal2020,Yeh2023,Dayal2024,Madau2024}. Understanding reionization is crucial for linking early-universe observations to the large scale structures in the present-day universe \citep[see][for reviews]{Barkana&Loeb2001,Zaroubi2013,Gnedin&Madau2022}.

Ionization fronts, the interfaces separating ionized and neutral regions, are the key features of the reionization process \citep{Shapiro2006, Whalen2008,Davies2016}. These fronts propagate as ionizing photons traverse the IGM, interacting with baryonic matter. The properties of ionization fronts - such as their velocity and shape - directly influence the morphology of reionization and may encode information about the distribution of ionizing sources, the density structure of the IGM, and the interplay of photoionization heating and cooling. Studying these properties provides a means to connect small-scale reionization physics to large-scale observational signatures, including the 21 cm signal and the thermal evolution of the IGM \citep{Furlanetto2006,Pritchard2012,McQuinn2016,Ghara2020}.

Ionization fronts also affect the thermal state and structure formation within the IGM. As these fronts pass through neutral regions, they deposit energy into the surrounding medium, heating the gas to tens of thousands of Kelvin. This heating process sets the thermal state of the post-reionization IGM \citep{D'Aloisio2019,Nasir&Daloisio2020}. Furthermore, by increasing the local Jeans mass, heating suppresses the formation of low-mass galaxies, introducing thermal feedback that influences cosmic structure formation during and after reionization \citep{Okamoto2008, Katz2020}.

Reionization timing maps provide a powerful approach to studying the physics of ionization front. These maps record when each region of the IGM transitions from neutral to ionized. The gradient of the timing field reveals the velocity of ionization fronts, while higher-order derivatives characterize their shape. Previous studies have used timing maps to investigate ionization front velocity distributions and their connection to post-ionization temperatures \citep{Deparis2019, D'Aloisio2019}. Additionally, timing maps have been used to explore the topology of ionized regions and the anisotropic growth of bubbles \citep{Thelie2023}.

In this work, we use the high-resolution CROC simulations, which couple radiative transfer with hydrodynamics, to generate detailed reionization timing maps. These simulations enable the characterization of ionization front velocities and shapes across cosmic time, offering insights into the evolution of reionization. By integrating simulation results with theoretical models, we aim to inform upcoming 21 cm experiments, such as the Hydrogen Epoch of Reionization Array (HERA) and the Square Kilometer Array (SKA), which will probe reionization with unprecedented sensitivity \citep{mellema2013,DeBoer2017,Ghara2017}.

This paper is organized as follows: Section~\ref{sec:methods:sim} describes the CROC simulations and the methodology for constructing reionization timing maps and deriving ionization front properties. Section~\ref{sec:methods:theory} introduces the analytical model based on the barrier-crossing formalism, which effectively describes the statistical properties of reionization. In Section~\ref{sec:results}, we present the velocity and curvature statistics of ionization fronts, compare simulation results to analytical predictions, and discuss the implications for reionization dynamics. Finally, Section~\ref{sec:discussion} summarizes the key findings.

%%%%%%%%%%%%%%%%%%%%%%%%%%%%%%%%%%%%%%%%
\section{Methodology: Simulation}\label{sec:methods:sim}

The Cosmic Reionization on Computers (CROC) project is one of the most sophisticated efforts to model cosmic reionization. CROC simulations solve coupled equations of radiative transfer, hydrodynamics, and chemical evolution, modeling the interaction between ionizing sources and the intergalactic medium (IGM). Physical processes such as star formation, stellar feedback, atomic and molecular cooling, and ionizing photon propagation are included to capture the complexity of reionization dynamics.

\subsection{Reionization Timing Maps}

The neutral hydrogen fraction ($x_{\rm HI}$) is tracked over time within each simulation cell. Reionization timing is defined as the moment when $x_{\rm HI}$ falls below a specified threshold (which in this work we choose to be $x_{\rm HI} = 0.1$). These timing maps provide a spatiotemporal record of ionization front evolution across the simulation volume.

\begin{figure}[htb!]
\centering
\includegraphics[width=\columnwidth]{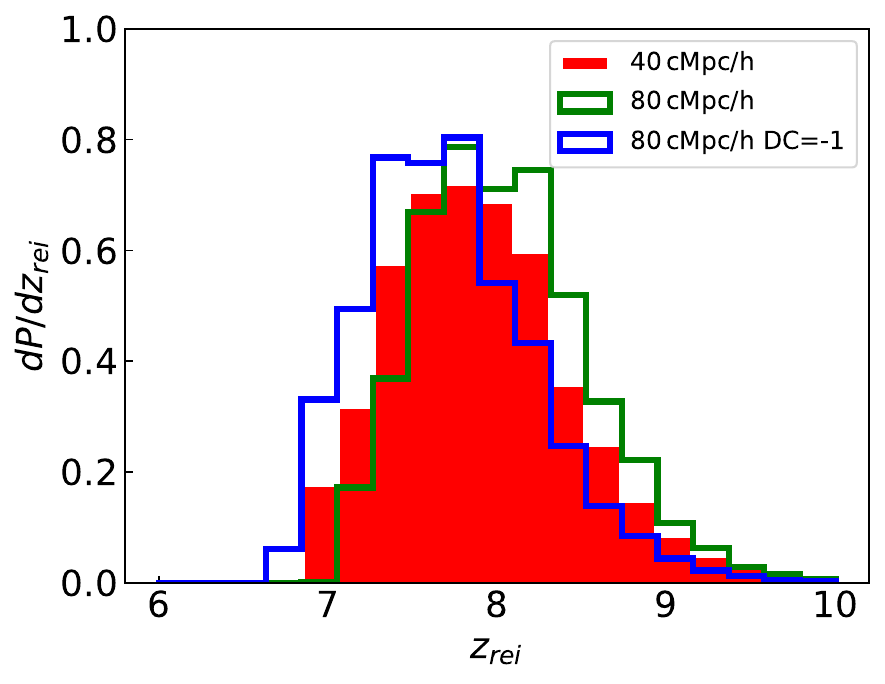}
\caption{PDF of reionization redshift ($z_{\rm rei}$) for simulations with different box sizes: 40 $h^{-1}$ Mpc, 80 $h^{-1}$ Mpc, and 80 $h^{-1}$ Mpc with a later reionization history (DC $= -1$). The distributions reflect spatial variations in reionization timing influenced by local density and large-scale structure.}
\label{fig:zrei_pdf}
\end{figure}

To examine the distribution of reionization timing, we compute the probability distribution function (PDF) of the redshifts at which reionization occurs ($z_{\rm rei}$). Figure~\ref{fig:zrei_pdf} shows the PDF for simulations with box sizes of 40 $h^{-1}$ Mpc, 80 $h^{-1}$ Mpc, and another 80 $h^{-1}$ Mpc box with a later reionization history. The later reionization history is achieved by introducing a ``DC mode" \citep{Sirko2005,Gnedin2011},  representing a deviation of the simulation volume's mean density from the true cosmic mean. This box, labeled as ``DC $=-1$", corresponds to a negative density contrast compared to the cosmic mean. The smaller simulation box reionizes later on average, while the negative density contrast further delays reionization. These patterns reflect the influence of local density variations and large-scale structure on the reionization process.

\begin{figure*}[htb!]
\centering
\includegraphics[width=\textwidth]{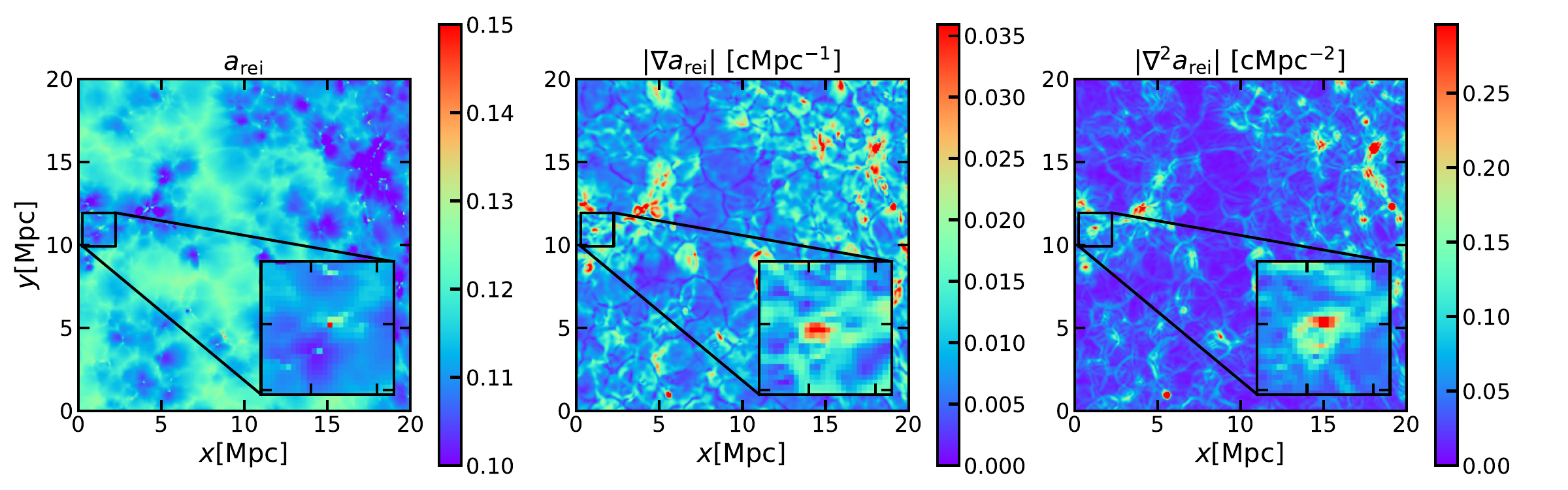}
\caption{Reionization timing field ($a_{\rm rei}$) and its derivatives in a 20 Mpc subvolume. Left: Timing field, where bluer regions reionize earlier. Middle: Gradient magnitude ($|\nabla a_{\rm rei}|$), which is inversely proportional to the speed of ionization front propagation. Right: Laplacian ($|\nabla^2 a_{\rm rei}|$), highlighting regions of rapid spatial variation in reionization timing.}
\label{fig:timing_maps}
\end{figure*}

Figure~\ref{fig:timing_maps} presents the reionization timing field in terms of the reionization scale factor ($a_{\rm rei}$) and its derivatives for a 20 Mpc subvolume of the simulation. The left panel shows the timing field, where earlier reionization appears as bluer regions. A zoomed-in view reveals small-scale variations in reionization timing, shaped by the density structure of the IGM and the spatial distribution of ionizing sources.

The middle panel displays the gradient magnitude ($|\nabla a_{\rm rei}|$), which is inversely proportional to the speed of ionization front propagation. Regions with lower $|\nabla a_{\rm rei}|$ are associated with faster-moving fronts, typically found in underdense regions. The right panel illustrates the Laplacian ($|\nabla^2 a_{\rm rei}|$), highlighting regions with rapid spatial variation in the timing field. These regions often coincide with slower-moving fronts, influenced by complex density structures and overlapping ionization fronts.

\subsection{Ionization Front Velocity}

The velocity of an ionization front is derived from the gradient of the reionization timing field. At a given location $\vec{x}$, the velocity is expressed as:
\begin{equation}
v_{I} = \frac{1}{\left|\nabla t_{\rm rei}\right|},
\end{equation}
where $\nabla t_{\rm rei}$ is the spatial gradient of the timing field (i.e. in units of time). This formulation directly relates the speed of ionization front propagation to the timing field.

Regions with identical reionization times can result in artificially high velocity estimates, necessitating careful consideration of the temporal resolution and the exclusion of outliers from the analysis.

\subsection{Ionization Front Curvature}

The curvature of an ionization front is calculated using the second-order derivatives of the reionization timing field. The Hessian matrix of second derivatives, $t_{{\rm rei},ij}$, is used to compute the projected curvature tensor:
\begin{equation}
C_{ij} = \frac{1}{|\nabla t_{\rm rei}|} P^k_i t_{{\rm rei},kl} P^l_j,
\end{equation}
where $P^k_i = \delta^k_i - n^k n_i$ is the projection tensor, and $\vec{n} = {\nabla t_{\rm rei}}/{|\nabla t_{\rm rei}|}$ is the unit normal vector to the ionization front. This projection confines the curvature calculation to the tangential plane of the front.

The two non-zero eigenvalues of $C_{ij}$ represent the principal curvatures, $k_1$ and $k_2$, of the ionization front. The Gaussian curvature is then defined as:
\begin{equation}
\mathcal{K} = k_1 k_2.
\end{equation}
This formalism captures the geometric properties of ionization fronts.

\section{Methodology: Analytical Model}\label{sec:methods:theory}

It is always helpful when numerical results can be interpreted with a simplified analytical model. One of the most enlightening (semi-)analytical descriptions of cosmic reionization was introduced by \citet{Furlanetto2004}. In this model, the cosmological ionization front is set by the condition
\begin{equation}
\delta(\vec{x},R_f) = b(t_{\rm rei},R_f), 
\label{eq:fm}
\end{equation}
where $\delta(\vec{x},R)$ is the Gaussian random field of initial conditions, smoothed with a filter of a spatial size $R$, $t_{\rm rei}(\vec{x})$ is the cosmic time when the location $\vec{x}$ is reionized, and a single function $b(t_{\rm rei},R)$ contains all the physics of reionization and is usually called a ``barrier". 

The primary limitation of the original \citet{Furlanetto2004} model is the assumption of an infinitely sharp barrier. Numerical simulations of reionization \citep{Kaurov2016,Gnedin2022} reveal that barriers are generally ``fuzzy," meaning they have a finite width that increases as the filter size $R$ decreases \citep[see Figure~11 in][]{Gnedin2022}. Despite this, a barrier $b(t_{\rm rei}, R)$ can always be constructed to match a given mean reionization history. When calibrated to a specific numerical simulation, such a model not only reproduces the mean reionization history but also reasonably captures the large-scale distribution of ionized regions \citep{Zahn2011,Kaurov2016}.

A particularly appealing feature of \citet{Furlanetto2004}-like modeling for this work is that the statistics of both the ionization front velocity and curvature can be computed in closed forms. The derivations are provided in Appendix~\ref{app:analytical}. We compare the analytical expressions to the simulation outputs below.

%%%%%%%%%%%%%%%%%%%%%%%%%%%%%%%%%%%%
\section{Results}\label{sec:results}
\subsection{Ionization Front Velocities}

\begin{figure}[th!]
\centering
\includegraphics[width=\columnwidth]{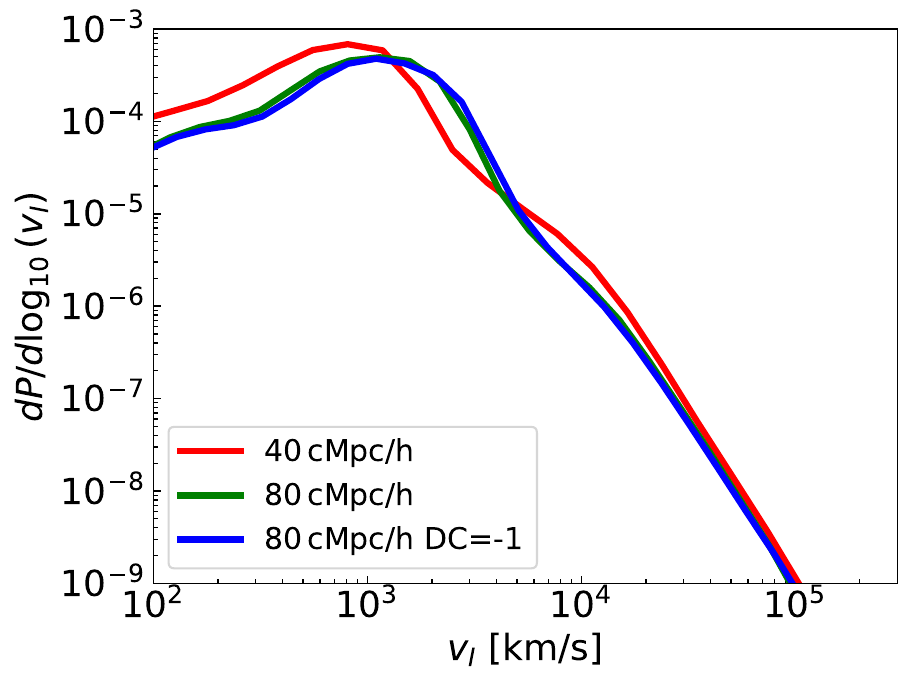}
\vspace{0.5cm}
\includegraphics[width=\columnwidth]{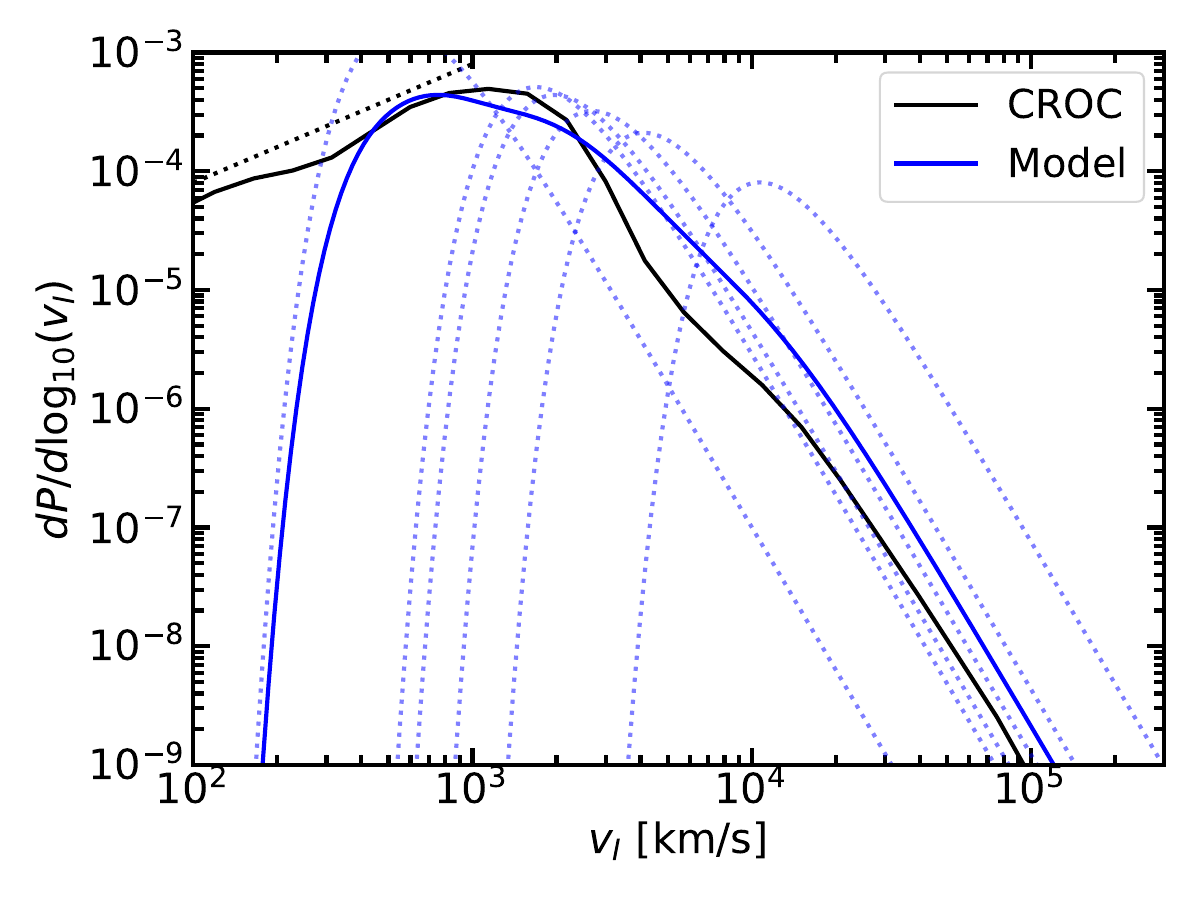}
\caption{ Ionization front velocity PDF from CROC simulations for three box sizes. The smaller box exhibits slightly lower peak velocities, while the 80 $h^{-1}$ Mpc simulations, including the DC = -1 case, show nearly identical velocity distributions, indicating that large-scale density variations have a limited effect on front velocities.  Bottom: Velocity PDF for the 80 $h^{-1}$ Mpc simulation (black, the same as the green line in the top panel) compared to the analytical model (blue) at $z=8$. Dotted lines represent the model predictions for fixed filtering scales ($R_f=0.001,0.003,0.01,0.03,0.1,0.3\times L_{\rm BOX}$), and the solid blue curve shows the velocity PDF integrated over $R_f$ with the weights given by the bubble size distribution. The dotted black line has a slope of 1.}
\label{fig:velocity_pdf}
\end{figure}

Figure~\ref{fig:velocity_pdf} compares the ionization front velocity distributions derived from CROC simulations and analytical models at $z=8$. The top panel shows velocity PDFs measured from simulations of three box sizes, consistent with those in Figure~\ref{fig:zrei_pdf}. The smaller 40 $h^{-1}$ Mpc box exhibits a slightly lower peak velocity, but the overall shapes of the velocity distributions are similar. While reionization timing varies significantly across these simulations, the velocities measured in the two 80 $h^{-1}$ Mpc boxes are nearly identical, indicating that box size and large-scale density variations have a limited effect on ionization front velocities.

The bottom panel compares the velocity PDF from the 80 $h^{-1}$ Mpc simulation (black) to predictions from the analytical model described in Section~\ref{sec:methods:theory} (blue). The analytical model assumes fixed filtering scales ($R_f$), depicted as dotted lines ranging from $0.03 L_{\rm BOX}$ to $0.3 L_{\rm BOX}$ uniformly spaced in log. The solid blue curve represents the PDF integrated over $R_f$ weighted by the bubble size distribution from Equation~9 of \protect\citet{Furlanetto2004}, under the assumption that each bubble contributes $4\pi R_f^2$ o the total velocity PDF. While the model captures the overall shape of the velocity PDF, it misses the low-velocity tail observed in the simulations. This is not surprising, as the exponential cutoff in the barrier-crossing model arises from the Gaussian distribution of density gradients (Equation~\ref{eq:pv_final}). In contrast, for a nonlinear density field, the PDF of densities follows a $\rho^{-2}$ scaling over a wide range of densities \citep[][Figure~1]{Klypin2018}. Replacing the Maxwellian distribution in Equation (\ref{eq:maxwellian_eta}) with a $\eta^{-2}$ scaling would produce a flat velocity PDF per unit $v_I$, or equivalently, a linearly rising PDF per unit $\log_{10}(v_I)$,which is approximately the behavior we observe in Figure~\ref{fig:velocity_pdf}.

\begin{figure}[th!]
\centering
\includegraphics[width=\columnwidth]{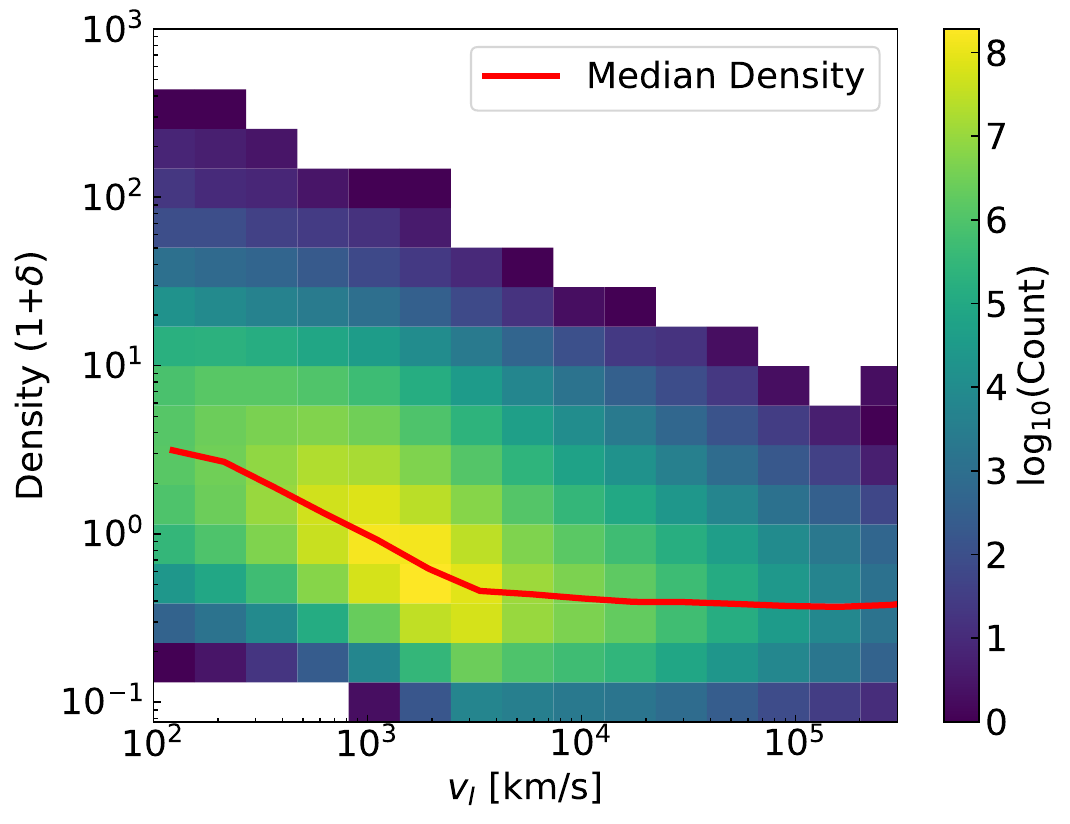}
\includegraphics[width=\columnwidth]{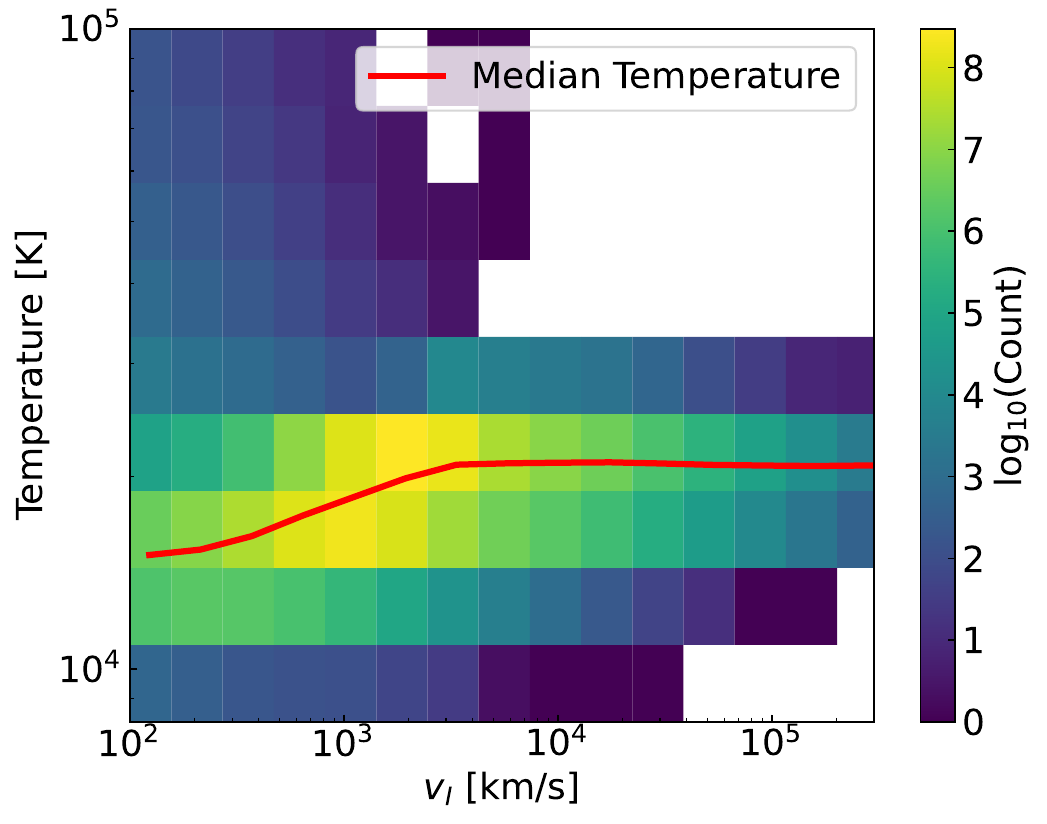}
\caption{Top: Relationship between ionization front velocity ($v_{I}$) and gas density ($\rho_{\rm rei}$), showing an inverse correlation where fronts move faster in underdense regions. Bottom: Relationship between $v_{I}$ and post-ionization gas temperature ($T$), with faster fronts producing higher temperatures, plateauing at $T \sim 2 \times 10^4$ K for $v_{I} > 3000$ km/s. Red lines indicate median trends.}
\label{fig:density_velocity}
\end{figure}

The velocity of ionization fronts is expected to depend on the local density of the IGM. In underdense regions, ionization fronts propagate quickly due to lower neutral hydrogen densities and reduced recombination rates. Conversely, overdense regions, such as filaments and dark matter halos, slow down ionization fronts because of higher recombination rates and greater photon consumption by the neutral media. The top panel of Figure~\ref{fig:density_velocity} demonstrates this relationship, showing the joint distribution of velocity and gas density. The median density (red line) decreases with increasing velocity, consistent with the anticipated inverse correlation. The scatter in the plot reflects the complexity of reionization, where small-scale density variations and local ionizing sources modulate ionization front velocities.

The propagation of ionization fronts also impacts the thermal state of the IGM. The bottom panel of Figure~\ref{fig:density_velocity} shows the relationship between ionization front velocity and gas temperature ($T$). The 2D histogram reveals that faster-moving fronts result in higher post-ionization temperatures, with the median temperature (red line) plateauing at $T \sim 2 \times 10^4$ K for $v_{I} > 3000$ km/s. These trends are consistent with the broader understanding of ionization front dynamics. \citet{D'Aloisio2019} also demonstrated that post-ionization temperatures are primarily determined by $v_{I}$, with minimal dependence on the ionizing radiation's spectral index. Their results indicate that slower fronts ($v_{I} \sim 50\text{--}2000$ km/s) produce temperatures of $T \sim 17,000\text{--}22,000$ K, while faster fronts near the overlap phase ($v_{I} \sim 10,000$ km/s) reach temperatures of $T \sim 25,000\text{--}30,000$ K. Additionally, \citet{Zeng2021} identified a velocity-dependent rise in post-ionization temperatures, where faster ionization fronts deposited more energy into the IGM.

\subsection{Evolution of Velocity Distribution During Reionization}

\begin{figure}[tbh!]
\centering
\includegraphics[width=\columnwidth]{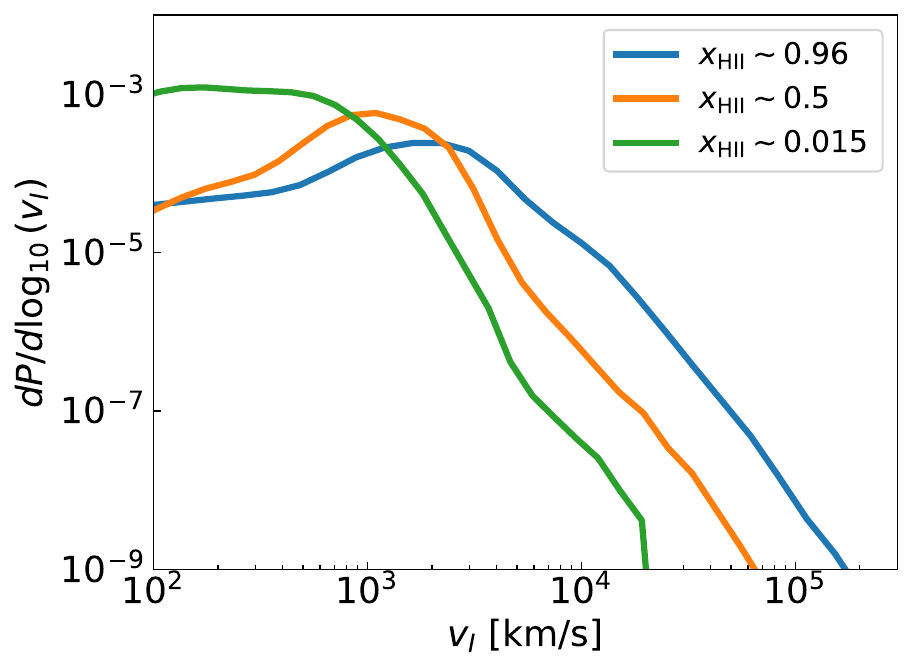}
\caption{Velocity distribution of ionization fronts ($v_{I}$) at various stages of reionization, characterized by the mass-weighted global ionized fraction ($x_{\rm HII})$. The distribution shifts from higher velocities in the early stages to lower velocities as reionization progresses.}
\label{fig:v_pdf_xHII}
\end{figure}

The velocity distribution of ionization fronts evolves throughout reionization, reflecting changes in the physical conditions of the IGM as the mass-weighted global ionized fraction ($x_{\rm HII}$) increases. Figure~\ref{fig:v_pdf_xHII} shows the velocity distribution ($v_{I}$) at different stages of reionization, characterized by varying $x_{\rm HII}$. In the early stages ($x_{\rm HII} \sim 0.015$), low-velocity fronts dominate due to the rapid ionized bubble growth in overdense regions near sources. As reionization progresses ($x_{\rm HII} \sim 0.5$), the velocity distribution broadens as ionization fronts propagate into underdense regions. In the later stages ($x_{\rm HII} \sim 0.96$), the distribution evolves further toward higher velocities, reflecting the accelerated growth of large ionized bubbles. This progression illustrates the transition from the slow initial expansion of ionized regions in dense areas to increasingly rapid propagation as reionization continues.

\subsection{Curvature}

\begin{figure*}[ht]
\centering
\includegraphics[width=\textwidth]{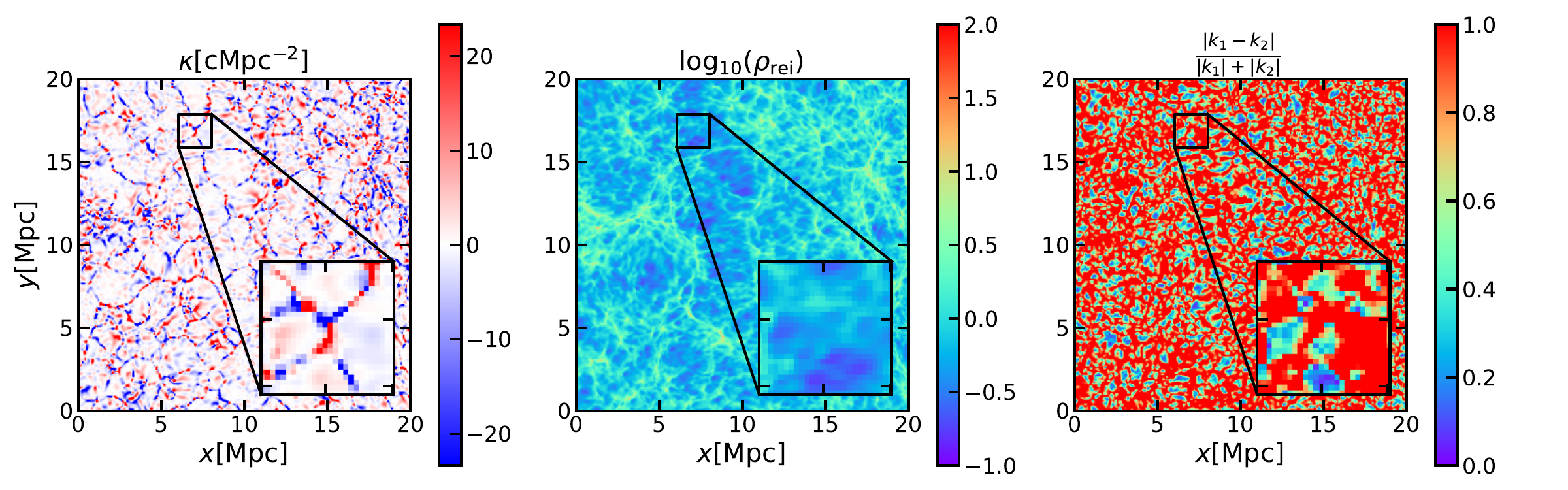}
\caption{Curvature field of the ionization front and its relation to density variations in a 20 Mpc subvolume. Left: Gaussian curvature $\kappa$, with blue and red regions indicating negative and positive curvatures, respectively, highlighting the non-sphericity of ionization fronts. Middle: Logarithmic local gas density ($\log_{10} \rho_{\rm rei}$), showing the density distribution. Right: Anisotropy metric $|k_1 - k_2| / (|k_1| + |k_2|)$, which quantifies deviations from isotropy. Insets zoom in on representative regions to emphasize small-scale variations in curvature and density.}
\label{fig:curvature}
\end{figure*}

\begin{figure}[th]
\centering
\includegraphics[width=\columnwidth]{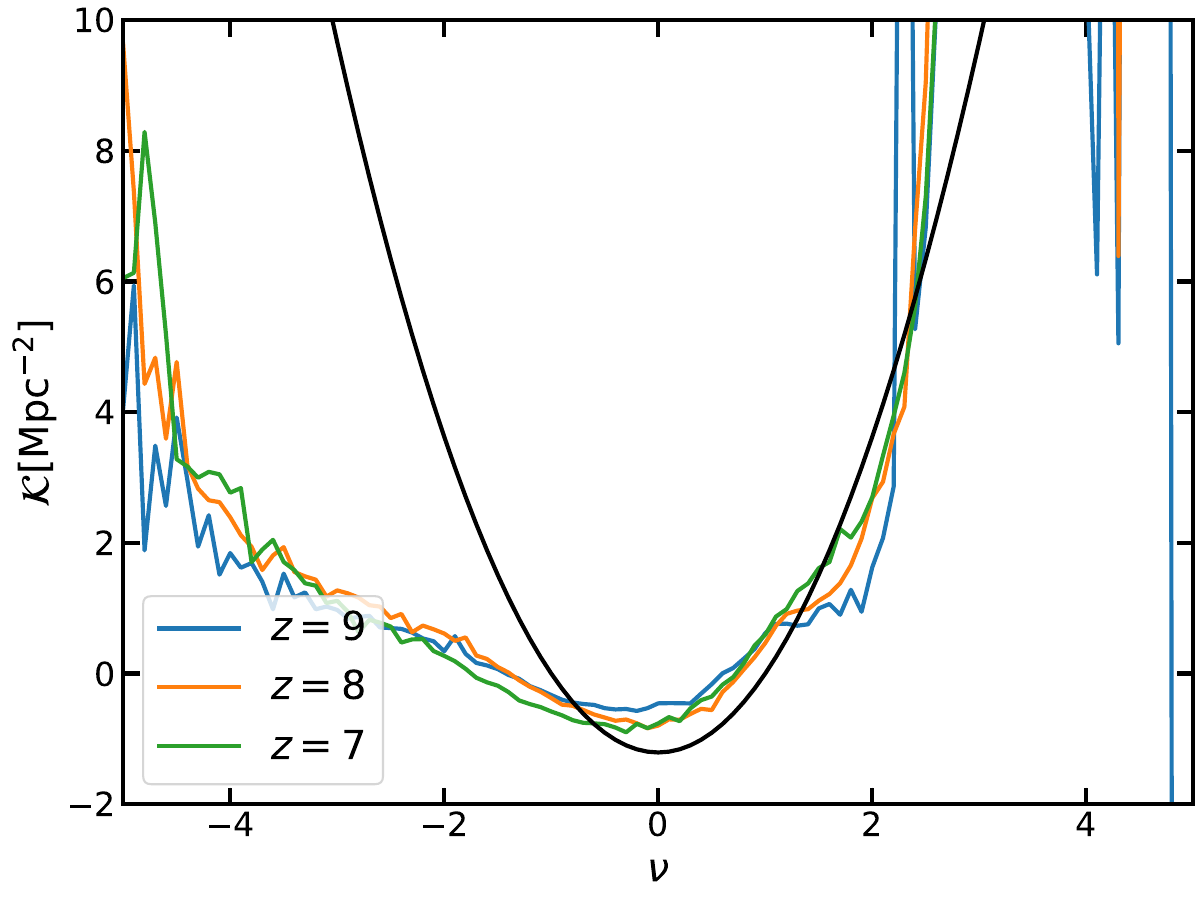}
\caption{The mean curvature at fixed density as a function of the normalized and scaled density, $\nu\equiv \delta/\sigma_0$, at several redshifts (colored lines). The same quantity $\langle\mathcal{K}\rangle$ computed in the \citet{Furlanetto2004} model in equation (\ref{eq:kavg}) is shown as a black line. The agreement for $\nu > 0$ confirms the model's validity in overdense regions, while deviations for $\nu < -1$ reveal its limitation in accurately modeling ionization fronts in voids.}
\label{fig:kavg}
\end{figure}

The left panel of Figure~\ref{fig:curvature} demonstrates that ionization bubbles in the CROC simulations deviate significantly from simple spherical geometries. The curvature field of the ionization front, derived from the simulation, highlights the complexity of these structures. The right panel provides further insight into the non-sphericity of the ionization fronts using the metric $|k_2 - k_1| / (|k_1| + |k_2|)$. Here, $k_1$ and $k_2$ represent the principal curvatures, describing the front's bending along orthogonal directions. Specifically, $k_1$ corresponds to the smallest curvature and $k_2$ to the largest. Positive curvatures denote outward convexity, while negative curvatures indicate inward concavity. For perfectly spherical ionization bubbles, $k_1$ and $k_2$ would be equal everywhere. The metric $|k_2 - k_1| / (|k_1| + |k_2|)$ quantifies deviations from sphericity, with values near zero indicating nearly isotropic fronts and larger values reflecting more anisotropic morphologies. The analysis confirms that non-sphericity is typical of ionization fronts.

Further analysis of the curvature map reveals that the curvature of the ionization front is not directly proportional to the local gas density. This is evident from a comparison of the left and middle panels of Figure~\ref{fig:curvature}, where the middle panel shows the logarithmic local gas density ($\log_{10}\rho_{\rm rei}$). Instead, we discover that the curvature correlates more strongly with properties of the Gaussian random field. Specifically, the mean curvature of the ionization front ($\langle\kappa\rangle$) correlates well with $\nu$, the density contrast normalized by the standard deviation of the Gaussian random field, as is shown in Figure \ref{fig:kavg}. For $\nu > 0$, the good agreement between the simulations and the analytical model suggests that ionization fronts in overdense regions are well-described by the barrier-crossing theory. For context, in spherical bubbles, the Gaussian curvature is inversely proportional to the square of the bubble radius $R$, with $\kappa = 1/R^2$. Thus, higher curvature values correspond to smaller bubbles. However, in underdense regions ($\nu < -1$), the measured mean curvature is systematically smaller than the analytical predictions. This discrepancy likely arises from a limitation of the barrier-crossing formalism. In the formalism, a remaining neutral island at the bottom of a void is modeled as an isodensity contour at some smoothing scale $R$. In reality, however, a void is more likely to be ionized by an ionization front sweeping across it in a direction determined by the distribution of ionizing sources, with no regard to the details of the density distribution within the void.

%%%%%%%%%%%%%%%%%%%%%%%%%%%%%%%%
\section{Summary and Discussion} \label{sec:discussion}
%%%%%%%%%%%%%%%%%%%%%%%%%%%%%%%%

This study presents two key findings. First, ionization front velocities and curvatures are closely connected to the density structure of the IGM. Velocities are inversely related to local density, with rapid propagation occurring in underdense regions and slower progress observed in overdense environments. Curvature, in contrast, is more strongly correlated with the normalized density contrast, $\nu$. In overdense regions ($\nu > 0$), the curvature statistics from simulations align well with predictions from analytical models based on barrier-crossing theory. However, significant deviations emerge in underdense regions ($\nu < -1$) due to the limitations of analytical models in low-density environments.

Second, the dynamics of ionization fronts play a significant role in determining the thermal state of the IGM. Faster-moving fronts generate higher post-ionization temperatures, with a plateau at $\sim 2 \times 10^4$ K for velocities exceeding 3000 km/s. These results are consistent with prior work \citep{D'Aloisio2019}, which shows slower fronts producing temperatures of $T \sim 17,000$--$22,000$ K and faster fronts during the overlap phase reaching $T \sim 30,000$ K. This temperature evolution reflects the interplay between photoionization heating and various cooling mechanisms.

The consistency between simulation results and analytical models in moderately high-density regions indicates that barrier-crossing theory effectively captures not only the large-scale features of reionization such as the overall ionization history and the size distribution for ionized bubbles, but also more nuanced details, such as the velocity distribution of ionization fronts and even their local curvatures (at least in an average sense).

\acknowledgments

We thank Steven Furlanetto for helpful comments and suggestions that improved the paper. This work was supported in part by the NASA Theoretical and Computational Astrophysics Network (TCAN) grant 80NSSC21K0271. This manuscript has also been co-authored by Fermi Research Alliance, LLC under Contract No. DE-AC02-07CH11359 with the United States Department of Energy. This work used resources of the Argonne Leadership Computing Facility, which is a DOE Office of Science User Facility supported under Contract DE-AC02-06CH11357. An award of computer time was provided by the Innovative and Novel Computational Impact on Theory and Experiment (INCITE) program. This research is also part of the Blue Waters sustained-petascale computing project, which is supported by the National Science Foundation (awards OCI-0725070 and ACI-1238993) and the state of Illinois. Blue Waters is a joint effort of the University of Illinois at Urbana-Champaign and its National Center for Supercomputing Applications. We also acknowledge the support by grant NSF PHY-2309135 to the Kavli Institute for Theoretical Physics (KITP) and the support from the University of Chicago’s Research Computing Center.

\appendix

\section{Analytical Expressions for Ionization Front Velocity and Curvature in a Barrier-Crossing Formalism}
\label{app:analytical}

\subsection{PDF of Ionization Front Velocity}

Differentiating $t_{\rm rei}(\vec{x})$ from Equation~(\ref{eq:fm}) with respect to the comoving position $x$ at fixed $R$ (using Einstein's notation for derivatives),
\begin{equation}
    t_{{\rm rei}, i} = \frac{1}{\dot{b}}\delta_{,i},
    \label{eq:treigrad}
\end{equation}
we obtain the ionization front velocity in proper units as
\begin{equation}
    v_I = a_{\rm rei}\frac{|\dot{b}|}{|\nabla\delta|}.
    \label{eq:vi}
\end{equation}

From now on we follow the notation from \citet{BBKS} (hereafter BBKS). Let $\vec{\eta} = \nabla \delta$ be the gradient of the field, and let $\eta = |\vec{\eta}|$ denote the gradient magnitude. In a Gaussian random field $\eta$ follows a Maxwellian distribution (BBKS Equation~A4):
\begin{equation}
    p(\eta)\, d\eta = \frac{6\sqrt{3}}{\sqrt{2\pi}\sigma_1^3} \eta^2 e^{-3\eta^2 / 2\sigma_1^2} d\eta,
    \label{eq:maxwellian_eta}
\end{equation}
where \(\sigma_1 \equiv \sqrt{\langle\eta^2\rangle}\) is the first spectral moment (BBKS Equation~4.6c).

Substituting $\eta$ with $a_{\rm rei}|\dot{b}|/v$ from Equation~(\ref{eq:vi}), we finally obtain:
\begin{equation}
    p(v) = \frac{2}{\sqrt{2\pi}} \frac{v_0^3}{v^4} \exp\left[ -\frac{(v_0 / v)^2}{2} \right],
    \label{eq:pv_final}
\end{equation}
where $v_0 = a_{\rm rei}|\dot{b}|\sqrt{3}/\sigma_1$. This PDF describes the velocity distribution of the ionization front in the barrier crossing formalism.

\subsection{Mean Ionization Front Curvature}

Differentiating Equation~(\ref{eq:treigrad}) once more, we obtain:
\begin{equation}
    t_{\rm rei, ij} = \frac{1}{\dot{b}}\delta_{,ij} - \frac{\ddot{b}}{\dot{b}^3}\delta_{,i}\delta_{,j}.
\end{equation}

The second fundamental form of the ionization front at location $\vec{x}$ is then the largest nontrivial minor of the projected, normalized second-derivative tensor:
\begin{equation}
    C_{ij} = \frac{1}{|\nabla t_{\rm rei}|} P^k_i t_{\rm rei, kl} P^l_j = \frac{1}{|\nabla \delta|} P^k_i \delta_{,kl} P^l_j,
\end{equation}
where $P^k_i = \delta^k_i - n^k n_i$ is the projection tensor on the ionization front and $\vec{n}$ is the unit vector in the direction of $\nabla t_{\rm rei}$:
\begin{equation}
    \vec{n} = \frac{\nabla t_{\rm rei}}{|\nabla t_{\rm rei}|} = \frac{\nabla \delta}{|\nabla \delta|}.
\end{equation}
Here, $\delta^k_i$ is the Kronecker delta. Note that the same symbol $\delta$ is used for the Gaussian random field, which is a potential source of confusion.

The eigenvalues of $C_{ij}$ are one zero (in the direction of the gradient) and two principal curvatures $k_1$ and $k_2$, whose product is the Gaussian curvature of the ionization front:
\begin{equation}
    \mathcal{K} = k_1 k_2.
\end{equation}

Following BBKS (Appendix~A therein), let us orient the coordinate system such that the $z$-direction aligns with the gradient of the Gaussian random field: $\vec{\eta} = \eta \vec{e}_z$. In this orientation, the Gaussian curvature is expressed as the determinant of the second fundamental form:
\begin{equation}
    \mathcal{K} = \frac{1}{\eta^2} (\zeta_{11} \zeta_{22} - \zeta_{12}^2),
\end{equation}
where $\zeta_{ij}$ are the components of the Hessian matrix of $\delta$. Substituting the BBKS variables $x, y, z$ (from their Equation~A2), the curvature becomes:
\begin{equation}
    \mathcal{K} = \frac{1}{\eta^2} \left(\frac{\sigma_2^2}{9}(x+3y+z)(x-2z) - \zeta_{12}^2\right),
\end{equation}
where $\sigma_2 \equiv \sqrt{\langle (\nabla \cdot \eta)^2 \rangle}$ is the second spectral moment.

It is convenient to introduce a new dimensionless variable
\begin{equation}
    u \equiv \frac{\zeta_{12}}{\sigma_2}, \quad s \equiv \frac{\eta}{\sigma_1}.
\end{equation}
With these, the Gaussian curvature simplifies to:
\begin{equation}
    \mathcal{K} = \frac{\sigma_2^2}{\sigma_1^2} \kappa, \quad \kappa \equiv \frac{q}{s^2},
\end{equation}
where $q = \frac{1}{9}(x+3y+z)(x-2z) - u^2$.

The joint PDF of $s$, $x$, $y$, $z$, and $u$ is given in BBKS (Equation~A4 in their paper):
\begin{equation}
    p(s,x,y,z,u) = \frac{90\sqrt{15}}{(2\pi)^{5/2}\sqrt{1-\gamma^2}} s^2 e^{-\hat{Q}},
\end{equation}
where
\begin{equation}
    2\hat{Q} = \frac{(x-x_*)^2}{1-\gamma^2} + 15y^2 + 5z^2 + 15u^2 + 3s^2.
\end{equation}

The average of the dimensionless curvature $\kappa$ can be computed analytically:
\begin{equation}
    \langle\kappa\rangle = \frac{90\sqrt{15}}{(2\pi)^{5/2}\sqrt{1-\gamma^2}} \int ds\, dx\, dy\, dz\, du\, q e^{-\hat{Q}} = \frac{\gamma^2}{3}\left(\nu^2-1\right),
\end{equation}
where $\nu$ is the value of the Gaussian field in units of its rms, $\nu\equiv \delta/\sigma_0$. For a peak in $\delta$, $\nu$ is often referred to as the ``peak height".

Thus, the average dimensional curvature is:
\begin{equation}
    \langle\mathcal{K}\rangle = \frac{\sigma_2^2}{\sigma_1^2} \langle\kappa\rangle = \frac{\sigma_1^2}{3\sigma_0^2}\left(\nu^2-1\right).
    \label{eq:kavg}
\end{equation}

\bibliographystyle{apj}
\bibliography{main}

\end{CJK*}
\end{document}